\documentclass[prd,nofootinbib,english]{revtex4}

\usepackage[letterpaper,top=2cm,bottom=2cm,left=2.35cm,right=2.35cm,marginparwidth=1.75cm]{geometry}

\usepackage{graphicx,float}
\usepackage{amsmath,amssymb,amsfonts}
\usepackage{upgreek}
\usepackage{mathrsfs}
\usepackage{epsfig,color}
\usepackage[thinlines]{easytable}
\usepackage{pdfpages}
\usepackage{array}
\usepackage{cancel}
\usepackage{mathtools}
\usepackage{accents}
\usepackage{subfigure}
\usepackage{enumitem}
\usepackage[dvipsnames]{xcolor}
\usepackage{hyperref}
\usepackage{ulem}

\hypersetup{
    colorlinks=true,
    linkcolor=blue,
    filecolor=magenta,
   citecolor=blue
}

\def\be{\begin{equation}}
\def\ee{\end{equation}}
\def\bea{\begin{eqnarray}}
\def\eea{\end{eqnarray}}

\def\mL{\mathcal{L}}
\def\mH{\mathcal{H}}
\def\mS{\mathcal{S}}
\def\mG{\mathcal{G}}
\def\mA{\mathcal{A}}

\def\mC{\mathcal{C}}

\def\mP{\mathcal{P}}

\def\mS{\mathcal{S}}

\def\mM{\mathcal{M}}

\def\zi{\mathrm{i}}
\def\ze{\mathrm{e}}

\def\zd{\mathrm{d}}

\def\zA{\mathrm{A}}

\def\zL{\mathrm{L}}
\def\zR{\mathrm{R}}

\def\vi{\mathfrak{i}}
\def\vj{\mathfrak{j}}
\def\va{\mathfrak{a}}

\def\vc{\mathfrak{c}}

\def\vv{\mathfrak{v}}
\def\vt{\mathfrak{t}}

\def\vC{\mathfrak{C}}

\def\vp{\mathfrak{p}}

\def\e{\epsilon}
\def\ce{\varepsilon}

\def\pA{\mathfrak{p}_{\mathrm{A}}}
\def\tl{\tilde{\lambda}}

\def\tB{\tilde{B}}
\def\tT{\tilde{T}}
\def\tvt{\tilde{\mathfrak{t}}}
\def\hmG{\hat{\mathcal{G}}}

\begin{document}
\hfill  USTC-ICTS/PCFT-23-12
\title{Parity violating scalar-tensor model in teleparallel gravity and its cosmological application}

\author{Haomin Rao}
\email{rhm137@mail.ustc.edu.cn}
\affiliation{School of Fundamental Physics and Mathematical Sciences, Hangzhou Institute for Advanced Study, UCAS, Hangzhou 310024, China}
\affiliation{University of Chinese Academy of Sciences, 100049 Beijing, China}

\author{Dehao Zhao}
\email{dhzhao@mail.ustc.edu.cn}
\affiliation{Interdisciplinary Center for Theoretical Study, University of Science and Technology of China, Hefei, Anhui 230026, China}
\affiliation{Peng Huanwu Center for Fundamental Theory, Hefei, Anhui 230026, China}

\begin{abstract}
The parity violating model based on teleparallel gravity is a competitive scheme for parity violating gravity,
which has been preliminary studied in the literature.
To further investigate the parity violating model in teleparallel gravity, in this paper, we construct all independent parity-odd terms that are quadratic in torsion tensor and coupled to a scalar field in a way without higher-order derivatives. Using these parity-odd terms, we formulate a general parity violating scalar-tensor model in teleparallel gravity and obtain its equations of motion. To explore potentially viable models within the general model, we investigate the cosmological application of a submodel of the general model in which terms above the second power of torsion are eliminated. We focus on analyzing cosmological perturbations and identify the conditions that preserve the parity violating signal of gravitational waves at linear order while avoiding the ghost instability.
\end{abstract}

\maketitle

\section{Introduction}\label{Introduction}
Stimulated by experimental advances in gravitational waves (GWs) \cite{ligo1,ligo2} and the cosmic microwave background radiation \cite{CMB1,CMB2},
parity violating (PV) gravities attracted lots of interests in recent years.
The most famous PV gravity model is the Chern-Simons (CS) gravity \cite{Jackiw:2003pm,Alexander:2009tp},
which modifies general relativity (GR) by a parity-odd topological term composed of curvature.
The CS gravity predicts the amplitude birefringence phenomenon of GWs, that is, the left- and right-handed GWs have different amplitudes.
However, this model suffers from the problem of ghost instability \cite{Dyda:2012rj} and
its further extensions within the framework of Riemannian geometry \cite{Crisostomi:2017ugk,Gao:2019liu,Zhao:2019xmm}
dose not fully circumvent this difficulty because ghost modes still appear at high energy scales, as shown in Ref.~\cite{Bartolo:2020gsh}.
It is very difficult to have a ghost-free PV gravity model within the framework of Riemannian geometry.
To seek the possibilities we may go beyond the Riemannian geometry.

Teleparallel gravity (TG) is one of the alternative formulation of gravity,
which identifies gravity as the spacetime torsion in stead of the curvature \cite{Aldrovandi:2013wha,Bahamonde:2021gfp}.
In the TG framework, there is a GR equivalent TG model call teleparallel equivalent of general relativity (TEGR) \cite{Maluf:2013gaa}
and this provides another way to modify the GR.
Along this way,  a PV gravity model within the TG framework
called Nieh-Yan modified teleparallel gravity (NYTG) was recently proposed in Ref.~\cite{Li:2020xjt,Li:2021wij}.
The NYTG model modifies TEGR by the coupling between an axion-like field $\phi$ and the Nieh-Yan density \cite{Nieh:1981ww}.
The Nieh-Yan density is a parity-odd topological term, so at the background with $\partial_{\mu}\phi\neq 0$,
the Nieh-Yan coupling term violates the parity spontaneously.
The NYTG model predicts velocity birefringence phenomenon of GWs, that is, the left- and right-handed GWs have different propagating velocities.
More importantly, through detailed studied on the cosmological perturbations,
it was shown in Refs.~\cite{Li:2020xjt,Li:2021wij} that the NYTG model is ghost-free.
The post-Newtonian, astronomical and other cosmological tests of the NYTG model can be found
in Ref.~\cite{Rao:2021azn,Qiao:2021fwi,Wu:2021ndf,Cai:2021uup,Li:2023fto}.
Other recent studies on PV gravities can be found in
Refs.~\cite{Gong:2021jgg,Hohmann:2022wrk,Tong:2022cdz,
Zhang:2022xmm,Zhu:2022dfq,Zhu:2022uoq,Filho:2022yrk,Qiao:2022mln,Cai:2022lec,Chen:2022wtz,Zhu:2023lhv,Feng:2023veu,Boudet:2022nub,Bombacigno:2022naf}.

However, the Nieh-Yan density is not the only parity-odd term composed only of torsion.
An extention of the NYTG model was considered in Ref.~\cite{Hohmann:2020dgy,Li:2022mti},
which includes all parity-odd terms that are quadratic in torsion and composed only of torsion.
But through detailed studied on the cosmological perturbations,
it was shown in Ref.~\cite{Li:2022mti} that the extended NYTG model suffers from the problem of ghost instability again,
unless it reduces to the NYTG model.
Although the NYTG model is a rare ghost-free PV gravity model,
it hides a scalar dynamical degree of freedom at linear perturbation level in flat universe \cite{Li:2020xjt,Li:2021wij}.
This makes it tricky to deal with scalar perturbations in flat universe,
especially when considering primordial fluctuations in the early universe \cite{Cai:2021uup,Li:2023fto}.
All these circumstances motivate us to explore other possibilities of PV model within the TG framework

To open up new ideas, let us briefly recall the case of modified gravity in Riemannian geometry.
In Riemannian geometry, in order to avoid the Ostrogradski instability \cite{Woodard:2006nt,Woodard:2015zca},
the Lovelock's theorem \cite{Lovelock:1971yv,Lovelock:1972vz} greatly restricts the form of gravity models composed only of curvature.
However, the scalar-tensor models \cite{Kobayashi:2011nu,Gleyzes:2014dya,Gleyzes:2014qga,Heisenberg:2018vsk,Sotiriou:2008rp}
provides a broader way to build colorful gravity models in Riemannian geometry, in which a scalar field is included as a part of gravity in addition to curvature.
Among them, the Horndeski model \cite{Horndeski:1974wa,Nicolis:2008in,Deffayet:2009mn,Kobayashi:2019hrl,Deffayet:2013lga}
is the most general scalar-tensor model leading to second-order field equations.
Similarly, scalar-tensor models can also be constructed within the TG framework \cite{Hohmann:2018vle,Hohmann:2018dqh,Hohmann:2018ijr},
in which a scalar field is included as a part of gravity in addition to torsion.
Along this way, the Horndeski analogue in the TG framework was proposed in Ref.~\cite{Bahamonde:2019shr},
which includes general parity-even terms that are quadratic in torsion tensor and non-minimally coupled to a scalar field.
More importantly, the idea of the scalar-tensor model provides a new window for us to explore more PV models within the TG framework.
In fact, within the framework of symmetric teleparallel gravity \cite{Nester:1998mp,BeltranJimenez:2017tkd},
where gravity is identified as non-metricity tensor rather than curvature or torsion,
the idea of scalar-tensor model has been applied to explore PV gravity models \cite{Li:2021mdp,Conroy:2019ibo}
and has achieved gratifying results \cite{Li:2022vtn}.
This prompts us to believe that such an idea is also feasible within the TG framework.

In this paper, we will investigate PV scalar-tensor models within the TG framework.
First, we find out all independent parity-odd terms that are quadratic in torsion tensor and non-minimally coupled to a scalar field.
To avoid Ostrogradski instability, higher-order derivatives are forbidden.
Then, we construct the most general PV scalar-tensor gravity model including all these parity-odd terms and obtain its equations of motions.
Next, we apply the PV scalar-tensor model to cosmology and mainly focus on the analysis of cosmological perturbations and their stability.
In order to facilitate the study of cosmological perturbations, in this process, we only consider the submodel of the general model
which removes all terms above the second power of torsion.
Through detailed investigations on the cosmological perturbations,
we will find out the conditions that make the submodel ghost-free
while preserving the PV signal of GWs at the linear perturbation level.

The present paper is organized as follows. In section \ref{TGtheory}, we will give a brief introduction to the teleparallel gravity.
In section \ref{PVSTTG}, after first introducing the torsion decomposition, we will find out all parity-odd terms we need and use them to construct a general PV scalar-tensor gravity model.  To simplify further analysis, in section \ref{Cosperturbation}, we consider a simple subcase of the general model and study its cosmological linear perturbations. In section \ref{Conclusion}, we will give a summary of this paper.

In this paper, we adopt the unit $8\pi G=1$ and the signature $(-,+,+,+)$.
The indices of interior space are denoted by $A,B,C,...=0, 1, 2, 3$ and $a, b, c,...=1, 2, 3$.
They are lowered and raised by the Minkowski metric $\eta_{AB}$.
The spacetime indices are denoted by $\mu, \nu, \rho,...=0, 1, 2, 3$ and $i, j, k,...=1, 2, 3$.
They are lowered and raised by the spacetime metric $g_{\mu\nu}$.
The volume element is denoted as $\ce_{\mu\nu\rho\sigma}=\sqrt{-g}\e_{\mu\nu\rho\sigma}$,
where $g$ is determinant of the metric, $\epsilon^{\mu\nu\rho\sigma}\equiv\e_{\mu\nu\rho\sigma}$ is antisymmetric symbol which satisfies
$\epsilon_{0ijk}=\epsilon_{ijk}$ and $ \epsilon_{123}=1$.
In addition, we distinguish the spacetime affine connection $\Gamma^{\rho}_{~\mu\nu}$
and its associated covariant derivative $\nabla$ from the Levi-Civita connection $\mathring{\Gamma}^{\rho}_{~\mu\nu}$
and its associated covariant derivative $\mathring{\nabla}$ respectively.

\section{Teleparallel gravity}\label{TGtheory}

The TG theory is formulated in a spacetime endowed with a metric $g_{\mu\nu}$ and an affine connection $\Gamma^{\rho}_{~\mu\nu}$,
which is curvature free and metric compatible,
\be\label{TGconstrain}
R_{\mu\nu\rho}{}^{\sigma}=2\partial_{[\nu}\Gamma^{\sigma}{}_{\mu]\rho}
+2\Gamma^{\sigma}{}_{[\nu|\lambda|}\Gamma^{\lambda}{}_{\mu]\rho}=0,~
\nabla_{\rho}g_{\mu\nu}=\partial_{\rho}g_{\mu\nu}
-2\Gamma^{\sigma}{}_{\rho(\mu}g_{\nu)\sigma}=0.
\ee
Without curvature and nonmetricity,
the gravity is identified with torsion $T^{\rho}_{~\mu\nu}=2\Gamma^{\rho}{}_{[\mu\nu]}$ in the TG theory.
Such a spacetime can also be described by the tetrad $e^{A}_{~\mu}$ and the spin connection $\omega^{A}_{~B\mu}$.
They relates the metric $g_{\mu\nu}$ and the affine connection $\Gamma^{\rho}_{~\mu\nu}$ through the following relations
\be\label{metrictetradrelation}
g_{\mu\nu}=\eta_{AB}e^{A}_{~\mu}e^{B}_{~\nu},~~
\Gamma^{\rho}_{~\mu\nu}=e_{A}^{~\,\rho}(\partial_{\mu}e^A_{~\nu}+\omega^A_{~B\mu}e^B_{~\nu}),
\ee
where $e_{A}^{~\mu}$ is the inverse of $e^{A}_{~\mu}$,
which satisfies $e^{A}_{~\mu}e_{A}^{~\nu}=\delta^{\mu}_{~\nu}$ and $e^{A}_{~\mu}e_{B}^{~\mu}=\delta^{A}_{~B}$.
In the language of tetrad and spin connections, the torsion tensor can be expressed as
\be\label{torsion tensor}
T^{\rho}_{~\mu\nu}=2e_{A}^{~\,\rho}(\partial_{[\mu}e^{A}_{~\nu]}+\omega^{A}_{~B[\mu}e^{B}_{~\nu]}).
\ee
The teleparallel constraints (\ref{TGconstrain}) indicate that the spin connection can be in general expressed as
\be\label{spinconnection}
\omega_{~B \mu}^{A}=(\Lambda^{-1})^{A}_{~C} \partial_{\mu} \Lambda_{~B}^{C},
\ee
where $\Lambda^{A}_{~B}$ is Lorentz matrix which is position dependent and satisfies the relation
$\eta_{AB}\Lambda^A_{~C}\Lambda^B_{~D}=\eta_{CD}$ at any spacetime point.
Therefore, the tetrad $e^{A}_{~\mu}$ and the Lorentz matrix $\Lambda^{A}_{~B}$ can be regarded as the basic variables of the TG theory.
In this way, the teleparallel constraints (\ref{TGconstrain}) are automatically satisfied.

The simplest TG model is the so-called teleparallel equivalent of general relativity (TEGR) model whose action is
\begin{eqnarray}\label{TEGR}
	S_{\text{TEGR}}=\frac{1}{2}\int \zd^4x ~{|e|}\, \mathbb{T}+S_{m}\equiv
	\frac{1}{2}\int \zd^4x~{|e|} \;\left(-\frac{1}{4}T_{\alpha\mu\nu}T^{\alpha\mu\nu}-
	\frac{1}{2}T_{\alpha\mu\nu}T^{\mu\alpha\nu}+T^{\lambda}T_{\lambda}\right)+S_{m},
\end{eqnarray}
where $T_{\lambda}=T^{\sigma}{}_{\lambda\sigma}$ is the torsion vector,
${|e|}=\sqrt{-g}$ is the determinant of the tetrad $e^{A}_{~\mu}$, $\mathbb{T}$ is the torsion scalar,
and other matter with the action $S_m$ is assumed to be minimally coupled to the metric. It can be proved that the TEGR action (\ref{TEGR}) is identical to the Einstein-Hilbert action up to a surface term
\begin{eqnarray}
		S_{\text{TEGR}}=\frac{1}{2}\int \zd^4x \sqrt{-g}\;  \left(\mathring{R}+2\mathring{\nabla}_{\mu}T^{\mu} \right) +S_{m},
\end{eqnarray}
where the curvature scalar $\mathring{R}$ is defined by the Levi-Civita connection $\mathring{\Gamma}^{\rho}_{~\mu\nu}$
and considered as being fully constructed from the metric.
Since the surface term in the action does not affect the equations of motion,
we say that the TEGR is equivalent to GR at the level of the equations of motion \cite{Maluf:2013gaa}.

The coincidence that the TEGR model is equivalent to GR provides another way to modify the GR,
which is to modify the TEGR model within the TG framework. Unlike the case where the curvature is the second-order derivative of the metric $g_{\mu\nu}$ in Riemannian geometry,
the torsion is only the first-order derivative of the basic variables $e^{A}_{~\mu}$ and $\Lambda^{A}_{~B}$ in the TG theory.
This makes gravity model in TG theory very easy to avoid the Ostrogradsky instability brought by higher-order derivatives \cite{Woodard:2006nt,Woodard:2015zca}.
All we need to do is to ensure that the Lagrangian does not include the derivative of the torsion.
A variety of modified TG models have emerged, such as the most studied modified TG model, i.e., the $f(\mathbb{T})$ model \cite{Ferraro:2006jd,Cai:2015emx}, which generalizes $\mathbb{T}$ in the action (\ref{TEGR}) to a smooth function $f(\mathbb{T})$, and the new GR \cite{Hayashi:1979qx,Bahamonde:2017wwk},
which modifies the coefficients of  $T_{\alpha\mu\nu}T^{\alpha\mu\nu}$, $T_{\alpha\mu\nu}T^{\mu\alpha\nu}$ and $T^{\lambda}T_{\lambda}$ in the action (\ref{TEGR}) as undetermined constants.

\section{Parity violating scalar-tensor model in teleparallel gravity}\label{PVSTTG}

In this section, we will construct the general parity violating scalar-tensor model which are quadratic in torsion tensor and contain arbitrary first-order derivatives of a scalar field. First, we briefly introduce the irreducible decomposition of torsion tensor.

\subsection{Irreducible decomposition of torsion}\label{Decomposition}

In order to better serve the construction of independent terms composed of torsion,
we review the irreducible decomposition of torsion in this subsection.
The torsion tensor $T^{\rho}_{~\mu\nu}$ can be decomposed into three irreducible (Lorentz group) parts as follows \cite{Hehl:1994ue,McCrea:1992wa}
\be\label{decomposition}
T^{\rho}_{~\mu\nu}=\frac{2}{3}\, \delta^{\rho}{}_{[\mu}\vv_{\nu]}+\ce^{\rho}_{~\mu\nu\sigma}\va^{\sigma}
+\frac{4}{3}\, \vt^{\rho}{}_{[\mu\nu]},
\ee
where the vector part $\vv_{\mu}$, the axial part $\va^{\nu}$ and the tensor part $\vt_{\mu\nu\rho}$ are respectively defined as
\bea
& &\vv_{\mu}=T^{\nu}_{~\nu\mu},\\
& &\va_{\mu}=\frac{1}{6}\ce_{\mu\nu\rho\sigma}T^{\nu\rho\sigma},\\
& &\vt_{\mu\nu\rho}=T_{(\mu\nu)\rho}+\frac{1}{3}\left(\vv_{(\mu}g_{\nu)\rho}-\vv_{\rho}g_{\mu\nu}\right),
\eea
where $\vv_{\mu}$ and $\va^{\mu}$ each have 4 independent components
and the tensor part $\vt_{\mu\nu\rho}$ satisfies
\begin{eqnarray}
	\vt_{[\mu\nu]\rho}=\vt_{(\mu\nu\rho)}=0, \quad \vt_{\mu\alpha}{}^{\alpha}=\vt^{\alpha}{}_{\mu\alpha}=\vt^{\alpha}{}_{\alpha\mu}=0,
\end{eqnarray}
so $\vt_{\mu\nu\rho}$ has only 16 independent components.

Splitting the 24 components of torsion into 4+4+16 independent components
makes it easier to construct independent parity-even and parity-odd terms composed of torsion,
which can constitute the Lagrangian of TG models.
For example, when a scalar field $\phi$ is introduced and higher-order derivatives are forbidden,
the independent linear torsion terms are only as follows
\be\label{lineartorsion}
I_{0}=\vv^{\mu}\phi_{\mu},~J_{0}=\va^{\mu}\phi_{\mu},
\ee
where $I_{0}$ is parity-even, $J_{0}$ is parity-odd, and $\phi_{\mu}\equiv\partial_{\mu}\phi$.
Note that $\vt^{\mu\nu\rho}\phi_{\mu}\phi_{\nu}\phi_{\rho}=0$ due to the symmetry of $\vt^{\mu\nu\rho}$ itself.
The idea of constructing independent terms is that the terms of types $\vv$, $\va$ and $\vt$ must be independent.
In the same way, when we consider the quadratic torsion terms,
the terms of types $\vv \vv$, $\vv \va$, $\vv \vt$, etc. are also independent.
For example, in the absence of scalar field, there are 3 independent parity-even terms which are quadratic in torsion tensor
\be\label{PCscalar}
T_{\text{vec}}=\vv_{\mu}\vv^{\mu},~T_{\text{axi}}=\va_{\mu}\va^{\mu},~T_{\text{ten}}=\vt_{\mu\nu\rho}\vt^{\mu\nu\rho}.
\ee
Once it is allowed to couple with the first-order derivatives of a scalar field,
there will be more independent parity-even terms which are quadratic in torsion tensor as follows \cite{Bahamonde:2019shr},
\bea\label{PCscalarphi}
& &\nonumber\quad~
I_{1}=I_{0}^{2},~I_{2}=J_{0}^{2},~I_{3}=\vt^{\mu}\vv_{\mu},~I_{4}=\tvt^{\mu}\va_{\mu},
\\ & &
I_{5}=\vt^{\mu\rho\sigma}\vt^{\nu}_{~\rho\sigma}\phi_{\mu}\phi_{\nu},~
I_{6}=\vt^{\rho\sigma\mu}\vt^{\nu}_{~\rho\sigma}\phi_{\mu}\phi_{\nu},~
I_{7}=\vt^{\mu}\vt_{\mu},
\eea
where we have defined
\begin{eqnarray}
	\vt^{\mu}=\vt^{\mu\rho\sigma}\phi_{\rho}\phi_{\sigma},\quad \tvt^{\rho\mu\nu}=\ce^{\mu\nu\alpha\beta}\vt^{\rho}{}_{\alpha\beta},\quad \tvt^{\mu}=\tvt^{\rho\sigma\mu}\phi_{\rho}\phi_{\sigma}.
\end{eqnarray}
The elements of the set $\{T_{\text{vec}}, T_{\text{axi}}, T_{\text{ten}}, I_{1}, I_{2}, I_{3}, I_{4}, I_{5}, I_{6}, I_{7}\}$
are the basic building blocks of most curent parity preserving TG models. In fact, a very general scalar-tensor TG model has been considered in Ref.~\cite{Bahamonde:2019shr}, which includes all the above parity-even terms.
The model is called Bahamonde-Dialektopoulos-Levi Said model 
and is the Horndeski analog within the TG framework.

As mentioned in the introduction, in this paper, we will find out all parity-odd terms which are quadratic in torsion and couple to the first-order derivatives of a scalar field. We will see in a moment that the torsion decomposition technique can be of great help in this matter.

\subsection{Parity-odd terms which are quadratic in the torsion tensor}\label{PVTerm}
In this subsection, we consider the independent parity-odd terms which are quadratic in the torsion tensor.
They are the building blocks for constructing the PV scalar-tensor model in the next subsection.

In the absence of scalar field, there are two independent parity-odd terms
\be\label{PVscalar01}
P_{1}=\vv^{\mu}\va_{\mu},~P_{2}=\tvt^{\rho\mu\nu}\vt_{\rho\mu\nu}.
\ee
Some literatures on TG model customarily adopt another set of independent terms as
\be\label{PVscalar02}
\mP_{1}=\frac{1}{2}\tT^{\rho\mu\nu}T_{\rho\mu\nu},~
\mP_{2}=T_{\mu}\tT^{\mu},
\ee
where $\tT^{\rho\mu\nu}=(1/2)\ce^{\mu\nu\alpha\beta}T^{\rho}_{~\mu\nu}$ and $\tT_{\mu}=\tT^{\sigma}_{~\mu\sigma}=3\va_{\mu}$.
Note that  since the curvature vanishes, $\mP_1$ is actually the Nieh-Yan density \cite{Nieh:1981ww}, which is a topological term with odd parity.
It can be verified that the following invertible relations hold
\be\label{PVrelation1}
P_{1}=-\frac{1}{3}\mP_{2},~P_{2}=\frac{9}{4}\mP_{1}-\frac{3}{2}\mP_{2}.
\ee
So the set $\{\mP_{1},\mP_{2}\}$ is  equivalent to the set $\{P_{1},P_{2}\}$.
The elements of the set $\{P_{1},P_{2}\}$ are the basic building blocks of almost all current PV models within the TG framework.

Next, we consider the case involving scalar field coupling. To avoid the Ostrogradsky instability, we only consider the coupling to the first derivative of the scalar field. From the analysis in Sec.~\ref{Decomposition}, we know that the terms of types $\vv \vv$, $\vv \va$, $\vv \vt$, $\va \va$, $\va \vt$, and  $\vt \vt$ are independent.
Among them, terms of types $\vv\vv$ and $\va\va$ cannot produce parity-odd terms,
so we only consider terms of types $\vv \va$, $\vv \vt$, $\va \vt$, and $\vt \vt$ respectively below. For type $\vv \va$, only one parity-odd term can be constructed
\be\label{typeva}
J_{1}=\vv^{\mu}\va^{\nu}\phi_{\mu}\phi_{\nu}=I_{0}J_{0}.
\ee For type $\va \vt$, two parity-odd terms can be constructed
\be\label{typeat}
J_{2}=\va_{\rho}\vt^{\rho\mu\nu}\phi_{\mu}\phi_{\nu},~~
\tilde{J}_{2}=\va_{\rho}\vt^{\mu\nu\rho}\phi_{\mu}\phi_{\nu}.
\ee
Since $\vt_{\mu\nu\rho}=-2\vt_{\rho(\mu\nu)}$, we have $\tilde{J}_{2}=-2J_{2}$, which means only $J_2$ is independent. For type $\vv \vt$, only one parity-odd term can be constructed
\be\label{typevt}
J_{3}=-\ce^{\mu\nu\rho\sigma}\vv_{\mu}\vt^{\lambda}{}_{\nu\rho}\phi_{\sigma}\phi_{\lambda}=\tvt^{\mu}\vv_{\mu}.
\ee For type $\vt \vt$, six parity-odd terms can be constructed
\bea\label{typett}
& &\nonumber
J_{4}=-\ce^{\mu\nu\rho\sigma}\vt^{\alpha}_{~\mu\nu}\vt^{\beta}_{~\rho\sigma}\phi_{\alpha}\phi_{\beta}=\hat{\vt}^{\mu\nu}\phi_{\mu}\phi_{\nu},~~\qquad
J_{5}=\ce^{\mu\nu\rho\sigma}\vt^{\lambda}_{~\mu\nu}\vt_{\lambda\alpha\rho}\phi_{\sigma}\phi^{\alpha}
=\tvt^{\lambda\mu\nu}\vt_{\lambda\rho\mu}\phi_{\nu}\phi^{\rho},
\\ & &\nonumber
J_{6}=\ce^{\mu\nu\rho\sigma}\vt^{\lambda}_{~\mu\nu}\vt_{\rho\alpha\lambda}\phi_{\sigma}\phi^{\alpha}
=\tvt^{\lambda\mu\nu}\vt_{\mu\rho\lambda}\phi_{\nu}\phi^{\rho},\quad
\tilde{J}_{56}=\ce^{\mu\nu\rho\sigma}\vt^{\lambda}_{~\mu\nu}\vt_{\rho\alpha\lambda}\phi_{\sigma}\phi^{\alpha}
=\tvt^{\lambda\mu\nu}\vt_{\mu\rho\lambda}\phi_{\nu}\phi^{\rho},
\\ & &
J_{7}=-\ce^{\mu\nu\rho\sigma}\vt_{\alpha\mu\nu}\vt_{\rho\beta\gamma}\phi_{\sigma}\phi^{\alpha}\phi^{\beta}\phi^{\gamma}=\tvt^{\mu}\vt_{\mu},\qquad
\tilde{J}_{7}=-\ce^{\mu\nu\rho\sigma}\vt_{\alpha\mu\nu}\vt_{\beta\gamma\rho}\phi_{\sigma}\phi^{\alpha}\phi^{\beta}\phi^{\gamma}.
\eea
It can be found that the six parity-odd terms in Eq.~(\ref{typett}) are not independent. Because the property of $\vt_{[\mu\nu]\rho}=\vt_{(\mu\nu\rho)}=0$ can lead to
$\vt_{\mu\nu\rho}+\vt_{\nu\rho\mu}+\vt_{\rho\mu\nu}=0$ and $\vt_{\mu\nu\rho}=-2\vt_{\rho(\mu\nu)}$, we can derive the relations
\begin{eqnarray}
	\tilde{J}_{7}=-2J_{7}, \quad \tilde{J}_{56}=-(J_{5}+J_{6}).
\end{eqnarray}
Meanwhile, it can be proved that the following identities hold
(see Appendix \ref{Appindentityproof} for proof)
\be\label{PVrelation2}
J_{5}=\frac{1}{6}(XP_{2}-2J_{4}),~J_{6}=\frac{1}{6}(XP_{2}+4J_{4}),~J_{7}=\frac{1}{6}XJ_{4},
\ee
where $X=-\phi^{\mu}\phi_{\mu}$.
Hence only two of the parity-odd terms of type $\vt\vt$ mentioned above are independent.
Therefore, we can conclude that in total there are only six independent parity-odd terms
which are quadratic in torsion tensor and contain at most the first-order derivatives of scalar field.
Considering that there is also a linear torsion term $J_0$,
here we can choose
\be\label{PVscalar1}
J_{0},~P_{1},~P_{2},~J_{1},~J_{2},~J_{3},~J_{4}.
\ee
as the independent basis of the parity-odd term. 
It should be clear that the choice of independent basis is not unique.
For example, we can introduce the following parity-odd terms
\be\label{PVscalar2}
\mP_{0}=\tilde{T}^{\mu}\phi_{\mu},~
\mP_{3}=-T^{\mu}\tilde{T}^{\nu}\phi_{\mu}\phi_{\nu},~
\mP_{4}=T^{\mu}_{~\nu\rho}\phi_{\mu}\phi^{\nu}\tilde{T}^{\rho},~
\mP_{5}=2\tilde{T}^{\mu\nu\rho}\phi_{\mu}\phi_{\nu}T_{\rho},~
\mP_{6}=\frac{1}{2}\tilde{T}^{\mu\rho\sigma}T^{\nu}_{~\rho\sigma}\phi_{\mu}\phi_{\nu}.
\ee
It can be proved that the following identities hold
\bea\label{PVrelation3}
& &\nonumber
J_{0}=\frac{1}{3}\mP_{0},~
J_{1}=\frac{1}{3}\mP_{3},~
J_{2}=\frac{1}{18}(X\mP_{2}-\mP_{3})-\frac{1}{6}\mP_{4},
\\ & &
J_{3}=\frac{1}{2}(X\mP_{2}-\mP_{3})-\frac{3}{4}\mP_{5},~
J_{4}=-\frac{1}{2}(X\mP_{2}-\mP_{3})+\frac{3}{2}\mP_{4}+\frac{3}{4}\mP_{5}+\frac{9}{4}\mP_{6}.
\eea
Combining the identities in Eqs.~(\ref{PVrelation1}) and (\ref{PVrelation3}),
it can be verified that the linear transformation between the set $\{\mP_{0},\mP_{1},\mP_{2},\mP_{3},\mP_{4},\mP_{5},\mP_{6}\}$
and the set $\{J_{0},P_{1},P_{2},J_{1},J_{2},J_{3},J_{4}\}$ is reversible.
Thus there is another independent basis equivalent to the basis (\ref{PVscalar1}) as
\be\label{PVscalar3}
\mP_{0},~\mP_{1},~\mP_{2},~\mP_{3},~\mP_{4},~\mP_{5},~\mP_{6}.
\ee
This basis can facilitate the analysis of cosmological perturbations in Sec.~\ref{Cosperturbation}.
We will use these two equivalent basis interchangeably below as needed.

\subsection{Parity violating scalar-tensor model}

Since we have obtained the building bricks of the PV scalar-tensor model in the previous subsection,
we can construct the general PV scalar-tensor model within the TG framework as
\be\label{model}
S=\int \zd^4 x~|e|\,\left[\, \frac{\mathbb{T}}{2}+\mG(\phi,X,J_{0},P_{1},P_{2},J_{1},J_{2},J_{3},J_{4})\, \right ]+S_{m},
\ee
where $\mG$ can be any smooth function of $\phi,X,J_{0},P_{1},P_{2},J_{1},J_{2},J_{3},J_{4}$. Since we are mainly concerned with the parity-odd terms in this paper,
we keep the parity-even terms in its simplest form, which is the form of the TEGR model.
The model (\ref{model}) is a very general model, and all previously studied PV models in TG are just special cases of it.
For example, when $\mG=\phi(c_{1}\mP_{1}+c_{2}\mP_{2})+\frac{1}{2}X-V(\phi)$, where $c_1$ and $c_2$ are constants, the action (\ref{model}) reduces to
the action of the extended NYTG model in Ref.~\cite{Hohmann:2020dgy,Li:2022mti}
\footnote{More general parity-even terms and more complicated form  of scalar field are also considered in Ref.~\cite{Hohmann:2020dgy}.}.
If further let $c_2=0$, the action will be reduced to the action of the NYTG model in Ref.~\cite{Li:2020xjt,Li:2021wij}.

This general PV scalar-tensor model has two kinds of gauge symmetries:
the diffeomorphism and the local Lorentz transformation, the later makes the following change:
\be\label{LT}
e^{A}_{~\mu}\rightarrow(L^{-1})^{A}_{~B}e^{B}_{~\mu},~ \Lambda^{A}_{~B}\rightarrow\Lambda^{A}_{~C}L^{C}_{~B},
\ee
where $L^{A}_{~B}$ is also the Lorentz matrix.
It's easy to prove that the metric $g_{\mu\nu}$ and the torsion tensor $T_{~\mu \nu}^{\rho}$
are invariant under the transformation (\ref{LT}), so is the action (\ref{model}).
Due to the local Lorentz invariance, we can always choose the gauge $\Lambda^{A}_{~B}=\delta^{A}_{~B}$, i.e., $\omega^{A}_{~B\mu}=0$.
This is the Weitzenb\"{o}ck connection which had been frequently adopted in the literature.
This gauge is also called the Weitzenb\"{o}ck gauge.

The equations of motion of the model (\ref{model}) follow from the variations with respect
to tetrad $e^{A}_{~\mu}$ and Lorentz matrix $\Lambda^{A}_{~B}$ separately:
\bea
\mathring{G}^{\mu\nu}+N^{\mu\nu}&=&\Theta^{\mu\nu},\label{eom1}\\
N^{[\mu\nu]}&=&0,\label{eom2}
\eea
where $\mathring{G}^{\mu\nu}$ is the Einstein tensor fully determined by the metric,
$\Theta^{\mu\nu}=(2/\sqrt{-g})(\delta S_m/\delta g_{\mu\nu})$ is the energy-momentum tensors for matters,
and
\be\label{nmn}
N^{\mu\nu}=\vC g^{\mu\nu}+\mS^{\mu\nu}+(\nabla_{\sigma}-\vv_{\sigma})\mA^{\mu\nu\sigma},
\ee
where we have defined
\bea
& &\nonumber
\vC=-\mG+\frac{1}{3}\mG_{0}J_{0}+(\frac{1}{3}\hmG_{1}+X\mG_{3})P_{1}+\hmG_{2}P_{2}
+(\frac{1}{3}\mG_{1}+\mG_{3})J_{1}
+(\frac{1}{3}\mG_{2}+6\mG_{4})J_{2}+\mG_{3}J_{3}+\mG_{4}J_{4},\quad\quad\quad\quad
\\& & \nonumber
\mS^{\mu\nu}=(3\mG_{3}P_{1}-2\mG_{X})\phi^{\mu}\phi^{\nu}
+\frac{1}{9}\left[(2\mG_{1}-3\mG_{3})I_{0}+2\mG_{5}\right]\left(3\phi^{(\mu}\va^{\nu)}+2\tvt^{(\mu\nu)\rho}\phi_{\rho}\right)
\\ & &\nonumber \quad\quad~
+(2\mG_{1}-3\mG_{2})J_{0}\phi^{(\mu}\vv^{\nu)}
+\frac{2}{9}(\mG_{2}-9\mG_{4})\left(3\va^{(\mu}\vt^{\nu)}-6\tau_{1}^{\mu\nu}-3\tau_{2}^{\mu\nu}+2\tvt^{(\mu\nu)\rho}\vt_{\rho}\right)
\\ & &\nonumber \quad\quad~
+\frac{1}{9}(2\hmG_{1}-3X\mG_{3})\left(2\tvt^{(\mu\nu)\rho}\vv_{\rho}+3\vv^{(\mu}\va^{\nu)}\right)
+\hmG_{2}\left(9\tvt^{(\mu\nu)\rho}\va_{\rho}-2\hat{\vt}^{\mu\nu}\right),
\\& & \nonumber
\mA^{\rho\mu\nu}=\frac{1}{3}(\mG_{2}-6\mG_{1})J_{0}\, g^{\rho[\mu}\phi^{\nu]}+\frac{1}{3}(\mG_{1}I_{0}+\mG_{5})\ce^{\mu\nu\rho\sigma}\phi_{\sigma}
-\mG_{3}(\tau_{3}^{\rho\mu\nu}-\tau_{3}^{[\mu\nu]\rho})+\mG_{2}\phi^{\rho}\phi^{[\mu}\va^{\nu]}-3\hmG_{2}\tvt^{\rho\mu\nu}
\\ & &\nonumber \quad\quad~
+\frac{1}{3}\hmG_{1}\ce^{\mu\nu\rho\sigma}\vv_{\sigma}+\frac{1}{3}(X\mG_{2}-6\hmG_{1})g^{\rho[\mu}\va^{\nu]}
+2(\mG_{4}-\mG_{3})g^{\rho[\mu}\tvt^{\nu]}
+\frac{1}{3}(\mG_{2}-9\mG_{4})\ce^{\mu\nu\rho\sigma}\vt_{\sigma}-3\mG_{4}\tau_{4}^{\rho\mu\nu},
\eea
and $\mG_{\phi}=\partial{\mG}/\partial{\phi}$, $\mG_{X}=\partial{\mG}/\partial{X}$, $\mG_{i}=\partial{\mG}/\partial{J_{i}}$ and $\hmG_{i}=\partial{\mG}/\partial{P_{i}}$,
in addition
\be\nonumber \tau_{1}^{\mu\nu}=\phi^{(\mu}\vt^{\nu)\rho\sigma}\phi_{\rho}\va_{\sigma},~ \tau_{2}^{\mu\nu}=\phi^{(\mu}\vt^{\nu)\rho\sigma}\phi_{\sigma}\va_{\rho},~ \tau_{3}^{\rho\mu\nu}=\phi^{\rho}\ce^{\mu\nu\alpha\beta}\phi_{\alpha}\vv_{\beta},~\tau_{4}^{\rho\mu\nu}=\phi^{\rho}\tvt^{\sigma\mu\nu}\phi_{\sigma}. \ee
Similar to most modified TG models, the equation of motion (\ref{eom2}) from the variation of $\Lambda^{A}_{~B}$ is not independent of Eq.~(\ref{eom1}).
This is reasonable since the Lorentz matrix $\Lambda^{A}_{~B}$ can always be set to the identity matrix by the gauge transformation (\ref{LT}).
A more detailed explanation can be found in Ref.~\cite{Li:2021wij}.
There is another equation following from the variation of the action (\ref{model}) with respect to $\phi$,
\be\label{eom3}
\mathring{\nabla}_{\mu}\Phi^{\mu}-\mG_{\phi}=0,
\ee
where
\be
\Phi^{\mu}=-2\mG_{X}\phi^{\mu}+\mG_{1}J_{0}\vv^{\mu}+(\mG_{0}+\mG_{1}I_{0})\va^{\mu}
-\mG_{2}\vt^{\mu\rho\sigma}\phi_{\rho}\va_{\sigma}+2\mG_{3}\tvt^{(\mu\nu)\rho}\phi_{\nu}\vv_{\rho}
+2\mG_{4}\hat{\vt}^{\mu\nu}\phi_{\nu}.
\ee
It can be verified that when $\mG=\frac{1}{2}X-V(\phi)$,
Eq.~(\ref{eom3}) reduces back to the familiar equation $\mathring{\square}\phi+V_{\phi}=0$,
where $\mathring{\square}=-g^{\mu\nu}\mathring{\nabla}_{\mu}\mathring{\nabla}_{\nu}$
and $V_{\phi}$ is the first derivative of the potential $V(\phi)$ to the scalar field $\phi$.

\subsection{Flat universe background}

As a preliminary exploration of the PV scalar-tensor model (\ref{model}), in this subsection,
we apply the model to flat universe and investigate the effect of the PV terms on the background.

In flat universe, the metric can be expressed in rectangular coordinate as
\be\label{FRWmetric}
\zd s^{2}=g_{\mu\nu}\zd x^{\mu}\zd x^{\nu}=a^{2}\left(\zd\eta^{2}-\delta_{ij} \zd x^{i} \zd x^{j}\right),
\ee
where $a=a(\eta)$ is the scale factor, $\eta$ is the conformal time.
Unlike the case of Riemannian geometry, in TG theory the  connection is still arbitrary to some extent even after the metric is determined.
For this reason, as suggested in Refs~\cite{Hohmann:2019nat,Hohmann:2020zre,Coley:2022qug}, we should additionally require that
the connection is also homogeneous and isotropic, that is,
\be\label{FRWconnection}
\mL_{\xi}\Gamma^{\rho}_{~\mu\nu}=\nabla_{\mu}\nabla_{\nu}\,\xi^{\rho}
-\nabla_{\mu}(T^{\rho}_{~\nu\sigma}\xi^{\sigma})=0,
\ee
where $\xi^{\mu}$ represents all Killing vector fields corresponding to the metric (\ref{FRWmetric}).
Combining Eqs.~(\ref{FRWmetric}) and (\ref{FRWconnection}) selected the flat universe solution
in which the tetrad $e^{A}_{~\mu}$ and spin connection $\omega^{A}_{~B\mu}$ have the following forms
\be\label{flatuniverse}
e^{A}_{~\mu}=a\delta^{A}_{~\mu}, ~\omega^{A}_{~B\mu}=0.
\ee
It can be verified that the background solution (\ref{flatuniverse}) leads to
\be\label{PVtermzero}
J_{0}=P_{1}=P_{2}=J_{1}=J_{2}=J_{3}=J_{4}=0,
\ee
thus $\mG=\mG(\phi,X)$ is just a function of the scalar field and its first-order derivative.
It seems that the PV terms has no effect on the background.
We can also examine this conclusion from the perspective of the equations of motion.
Putting the solution (\ref{flatuniverse}) into Eqs.~(\ref{eom1}) and (\ref{eom3}), we obtain the background equations as
\bea
& &\label{flateom1} 3\mH^{2}=a^{2}\rho+2\mG_{X}{\phi'}^{2}-a^{2}\mG,\\
& &\label{flateom2} 2\mH'+\mH^{2}=-a^{2}(p+\mG),\\
& &\label{flateom3}  2\mG\phi''+(2\mG_{X}'+4\mG_{X}\mH)\phi'-a^{2}\mG_{\phi}=0,
\eea
where $\mH=a'/a$ is conformal Hubble rate, prime represents the derivative with respect to the conformal time $\eta$, and $\rho$ and $p$ denote the energy density and pressure of other matter. The background equations (\ref{flateom1})-(\ref{flateom3}) are exactly the same as the case where the Lagrangian of the scalar field is $\mG(\phi, X)$ in GR. This clearly confirms that the PV terms has no effect on the flat universe background.
If we want to probe the PV signals of the model (\ref{model}) in flat universe, we need to investigate its cosmological perturbations.

\section{Cosmological perturbations and stability analysis}\label{Cosperturbation}

To analyze the PV signals and stability of the model (\ref{model}), we investigate the cosmological perturbations of the model (\ref{model}) around the flat universe background in this section.
We will focus on the quadratic action of perturbations, find out the necessary conditions to make the model stable,
and explore the PV signals in GWs.

For the sake of simplicity, in the following we discard all terms above the second power of torsion
and take the parity-even part of the action (\ref{model}) as the simplest standard form.
This simplification allows the function $\mG$ to be reduced to
\be
\mG(\phi, X, \mP_{0},\mP_{1},\mP_{2},\mP_{3},\mP_{4},\mP_{5},\mP_{6})=\frac{1}{2}X-V(\phi)+\sum_{\vi=0}^{6}f_{\vi}(\phi,X)\mP_{\vi}.
\ee
where $f_{\vi}$ can be any smooth function of $\phi$ and $X$.
In addition, for the sake of convenience,
we also denote
\begin{eqnarray}
	f_{\vi \phi}\equiv\partial f_{\vi}/\partial \phi,\quad f_{\vi X}\equiv\partial f_{\vi}/\partial X, \quad f_{\vi \vj}\equiv f_{\vi}+Xf_{\vj},\quad f_{\vi \vj X}\equiv\partial f_{\vi \vj}/\partial X .
\end{eqnarray}
In this section, we will focus on the quadratic action of perturbations, which can be regarded as the effective action of the linear perturbations \cite{Mukhanov:1990me}. When applying the model to the inflation epoch, in that case the scalar field $\phi$ may be considered as the inflaton, we need to quantize these perturbations to have a mechanism for generating the primordial perturbations which seed the large scale structure at later time. For this purpose, the quadratic actions are indispensable. In the inflation epoch, we can ignore other matters except inflaton, so that in the following contents, we always choose $S_{m}=0$.
With the above simplifications, the action (\ref{model}) can be reduced to
\be\label{model0}
S=\int \zd^4 x~|e|\, \left[\, \frac{\mathbb{T}}{2}+\frac{1}{2}X-V(\phi)+\sum_{\vi=0}^{6}f_{\vi}(\phi,X)\mP_{\vi}\right],
\ee

From now on, we apply the model (\ref{model0}) to cosmology. From Eqs.~(\ref{flateom1})-(\ref{flateom3}), the background equations of the model (\ref{model0}) can be obtained as
\bea
& &\label{flateom01} 3 \mH^{2}=(1/2){\phi'}^{2}+a^{2}V,\\
& &\label{flateom02} 2 \mH'+\mH^{2}=-(1/2){\phi'}^{2}+a^{2}V,\\
& &\label{flateom03} \phi''+2 \mH \phi' +a^{2} V_\phi=0,
\eea
which are exactly the same as in GR, as expected.

After clearing the background, let us look at the perturbations.
We use the following parametrization for perturbed tetrad \cite{Izumi:2012qj,Golovnev:2018wbh}:
\bea\label{tetradperturbation}
& & e^{0}_{\ 0}=a(1+A),~\nonumber e^{0}_{\ i}=a(\beta_{,i}+\beta_{i}^{V}),
~\nonumber e^{c}_{\ 0}=a\, \delta^{c}_{\, i}\, (\chi_{,i}+\chi_{i}^{V}),\nonumber\\
& & e^{c}_{\ i}=a\, \delta^{c}_{\, j}\Big[ (1-\psi)\delta_{ij}+\alpha_{,ij}+\alpha_{j,i}^{V}-
              \epsilon_{ijk}(\lambda_{,k}+\lambda_{k}^{V})+\frac{1}{2}h^{T}_{ij}\Big],
\eea
the subscript $``,i"$ means $\partial_{i}$.
So the perturbed metric components have the familiar form:
\bea\label{metricperturbation}
& &g_{00}=-a^{2}(1+2A),~ g_{0i}=a^{2}(B_{,i}+B_{i}^{V}),\nonumber\\
& &g_{ij}=a^{2}[(1-2\psi)\delta_{ij}+2\alpha_{,ij}+\alpha_{i,j}^{V}+\alpha_{j,i}^{V}+h^{T}_{ij}],
\eea
where $B=\chi-\beta$ and $B^{V}_{i}=\chi^{V}_{i}-\beta^{V}_{i}$.
All the vector perturbations are transverse and denoted by the superscript $V$, both the
tensor perturbations are transverse and traceless and denoted by the superscript $T$.
In addition, the scalar field $\phi$ is decomposed as $\phi(\eta, \vec{x})=\bar{\phi}(\eta)+\delta\phi(\eta, \vec{x})$.

Due to the diffeomorphism invariance and the local Lorentz invariance,
it is safe to take the unitary gauge $\delta\phi=0,~\alpha=0,~\alpha_{i}^{V}=0$ and the Weitzenb\"{o}ck gauge $\omega^{A}_{~B\mu}=0$
at the same time on the general cosmological background with $\bar{\phi}'\neq 0$ \cite{Li:2021wij}
\footnote{On the background with $\bar{\phi}'=0$, $\delta\phi$ is gauge invariant, it is impossible to set $\delta\phi$ to zero by gauge transformation.
Therefore, there is no unitary gauge on the background with $\bar{\phi}'=0$. When we consider the case of Minkowski or de Sitter background, we should adopt other gauge. }
. Therefore, in the following contents, we will adopt the gauge
\begin{eqnarray}
	\delta\phi=0,~\alpha=0,~\alpha_{i}^{V}=0, ~\omega^{A}_{~B\mu}=0
\end{eqnarray}
to simplify our calculations,
We also introduce the gauge invariant scalar perturbation
\begin{eqnarray}
	\zeta=-(\psi+\mH\delta\phi/\phi')
\end{eqnarray}
representing the curvature perturbation of the hypersurfaces of constant $\phi$ field.
Then, we can choose $\zeta$, $A$, $B$, $\beta$, $\lambda$, $B_{i}^{V}$, $\beta_{i}^{V}$, $\lambda_{i}^{V}$ and $h^{T}_{ij}$ as independent variables.

Note that higher-order derivatives are just one of the sources of ghost modes.
Hybrid kinetic terms such as $x'y'=(1/2)[(x'+y')^2-(x'-y')^2]\equiv(1/2)(z_1'^2-z_2'^2)$ can also bring ghost modes.
Although there is no higher-order derivative in the model (\ref{model0}),
the parity-odd terms $\mP_{\vi}$ may lead to hybrid kinetic terms.
Therefore the model (\ref{model0}) is likely to suffer from the problem of ghost instability.
We will investigate the quadratic actions for the scalar, vector, and tensor perturbations of the model (\ref{model0}) separately in the following subsections.

\subsection{Quadratic action for scalar perturbations}\label{Sperturbation}

For scalar perturbations, we introduce notations $\tB=B_{, ii}$ and $\tl=\lambda_{,ii}$, and expand all scalar perturbations as follows
\be\label{Sexpand}
\zeta(\eta, \vec{x})=\int \frac{\zd^{3}k}{(2\pi)^{\frac{3}{2}}}\,
\zeta(\eta, \vec{k})\,\ze^{\zi\vec{k}\cdot\vec{x}}.
\ee
Then the quadratic action for the scalar perturbations can be directly obtained as
\bea\label{Saction1}
& &\nonumber
S^{(2)}_{S}=\int \zd\eta\, \zd^{3}k\, a^{2}\Big[
-3{\zeta'}^{2}+k^{2}(\zeta^{2}+2A\zeta)+2(\zeta'-\mH A)\tilde{B}
+6\mH\zeta'A-a^{2}VA^{2}
\\ & &\quad\quad\quad\quad\quad
+2f_{24}(A-\beta')\tl'
+(2f_{23}+4Xf_{3})\zeta'\tl
+2f_{23}(k^{2}\beta-\tilde{B})\tl
-4\vc_{1} A\tl+4\vc_{2} \zeta\tl
\Big],
\eea
where
\begin{align}
	\nonumber\vc_{1}&=X(\phi'f_{0X}+2\mH f_{1X}+3\mH f_{23X}),\\
	\nonumber\vc_{2}&=\phi'f_{0}-f_{1}'-f_{2}'+\mH(f_{23}+2Xf_{3}),
\end{align}
and we have simply marked $A^{*}B$ as $AB$, $A^{*}A$ as $A^{2}$, and so on.
It can be seen that $A$, $B$ are non-dynamical fields
and the variations of the action (\ref{Saction1}) with them lead to the following constraints
\bea
& & \label{Sceq1}
\mH A-\zeta'+f_{23}\tilde{\lambda}=0,
\\ & & \label{Sceq2}
-3\mH\zeta'-k^{2}\zeta+a^{2}VA+\mH\tB+2\vc_{1}\tl-f_{24}\tl'=0.
\eea
These constraint equations are just used to solve the non-dynamical variables $A$ and $B$.
One can eliminate these two non-dynamical variables from the action (\ref{Saction1}) by substituting
the constraints (\ref{Sceq1}) and (\ref{Sceq2}) back into it. After that,
the quadratic action for scalar perturbations can be expressed as
\bea\label{Saction2}
S^{(2)}_{S}=\int \zd\eta\, \zd^{3}k \bigg\{
\frac{1}{2}z^{2}(\zeta^{'2}-k^{2}\zeta^{2})+a^{2}\Big[2f_{24}\tl' (\zeta'/\mH-\beta')
+{\mC}_{1}\zeta'\tl+2k^{2}f_{23}\tl\beta+{\mC}_{2}\zeta\tl+{\mC}_{3}\tl^{2}\Big]
\bigg\},
\eea
where $z^{2}=a^{2}{\phi'}^{2}/\mH^{2}$, and
\bea
& &\nonumber
{\mC}_{1}=4Xf_{3}+\big(2a^{2}V\mH^{-2}-4\big)f_{23}-4\mH^{-1}\vc_{1},\\
&&\nonumber
{\mC}_{2}=4\vc_{2}-2k^{2}\mH^{-1}f_{23},
\\ & &\nonumber
{\mC}_{3}=4\mH^{-1}f_{23}\vc_{1}-a^{2}V\mH^{-2} f_{23}^{2}+a^{-2}\big[a^{2}\mH^{-1}(f_{23}f_{24})\big]'.
\eea
Obviously whether $f_{24}$ is zero or not will affect the number of dynamical degrees of freedom (DoFs) in the action (\ref{Saction2}),
so we will discuss it case by case below.

\subsubsection{The case of $f_{24}\neq0$}\label{Sperturbation01}

In the more general case with $f_{24}\neq0$, the kinetic terms $\tl'\beta'$ and $\tl'\zeta'$
generally contain ghost modes, as mentioned in the paragraph just before this subsection.
In order to explicitly see how many dynamical DoFs and how many ghost modes there are,
we define new perturbation variables $\gamma_{1}, \gamma_{2}$ in terms of the old variables $\beta$ and $\lambda$ as follows
\be\label{newscalar}
\gamma_{1}=\frac{1}{2}(\tilde{\lambda}-\beta+\mH^{-1}\zeta),~~
\gamma_{2}=\frac{1}{2}(\tilde{\lambda}+\beta-\mH^{-1}\zeta).
\ee
It can be verified that the transformation (\ref{newscalar}) is linearly reversible.
Then we can express the action (\ref{Saction2}) in terms of the new perturbation variables as
\be\label{Saction3}
S^{(2)}_{S}=\int \zd\eta\, \zd^{3}k\, \bigg\{
\frac{1}{2}z^{2}({\zeta'}^{2}-k^{2}\zeta^{2})+a^{2}\Big[2f_{24}({\gamma_{1}'}^{2}-{\gamma_{2}'}^{2})
+{\mC}_{4}\zeta'(\gamma_{1}+\gamma_{2})+U(\zeta, \gamma_{1}, \gamma_{2})\Big]\bigg\},
\ee
where
\bea
 & &\mC_{4}=(4-2a^{2}V\mH^{-2})f_{24}+\mC_{1},\nonumber\\
 &&\mC_{5}=4\vc_{2}+a^{-2}\big[(z^{2}-2a^{2})f_{24}\big]'.\nonumber\\
& & \nonumber
U(\zeta, \gamma_{1}, \gamma_{2})={\mC}_{3}(\gamma_{1}+\gamma_{2})^{2}+2k^{2}f_{23}(\gamma_{2}^{2}-\gamma_{1}^{2})+{\mC}_{5}\zeta(\gamma_{1}+\gamma_{2}).
\eea
The action (\ref{Saction3}) clearly shows that $\zeta$, $\gamma_{1}$ and $\gamma_{2}$ are all dynamical DoFs,
and one of $\gamma_{1}$ and $\gamma_{2}$ must be a ghost mode,
because the signs of the kinetic terms of $\gamma_{1}$ and $\gamma_{2}$ are always opposite.
This will cause the vacuum instability.
The only way for scalar perturbations to avoid ghost instability is to keep $f_{24}=0$.

\subsubsection{The case of $f_{24}=0$ on the background with $\phi'\neq0$}\label{Sperturbation02}
In the case of $f_{24}=0$, the quadratic action (\ref{Saction2}) reduces to the following one
\bea\label{Saction4}
S^{(2)}_{S}=\int \zd\eta\, \zd^{3}k \bigg\{
\frac{1}{2}z^{2}(\zeta^{'2}-k^{2}\zeta^{2})+a^{2}\Big[
{\mC}_{1}\zeta'\tl+2k^{2}X(f_{3}-f_{4})\tl\beta+{\mC}_{2}\zeta\tl+{\mC}_{3}\tl^{2}
\Big]\bigg\}.
\eea
It can be seen that $\beta$, $\tl$ are also non-dynamical fields.
On the background with $\bar{\phi}'\neq0$, the variations of the action (\ref{Saction4}) with respect to $\beta$ and $\tl$ lead to the following constraints
\bea\label{Sceq3}
& & (f_{3}-f_{4})\tilde{\lambda}=0,
\\ & &
{\mC}_{1}\zeta'+{\mC}_{2}\zeta+2k^{2}X(f_{3}-f_{4})\beta+2{\mC}_{3}\tl=0.\label{Sceq4}
\eea

If $f_{3}\neq f_{4}$, these constraint equations are just used to solve the non-dynamical variables $\tl$ and $\beta$.
One can eliminate these two non-dynamical variables from the action (\ref{Saction4}) by substituting
the constraints (\ref{Sceq3}) and (\ref{Sceq4}) back into it.
After that, the quadratic action for scalar perturbations can be reduced to
\bea\label{Saction5}
S^{(2)}_{S}=\int \zd\eta\, \zd^{3}k \left[\frac{1}{2}z^{2}(\zeta^{'2}-k^{2}\zeta^{2})\right].
\eea
It can be seen that the quadratic action (\ref{Saction5}) is exactly the same as that in GR.
There is only one scalar dynamical DoF and that DoF is healthy.

If $f_{3}= f_{4}$, then ${\mC}_{3}=0$ automatically, so the constraints (\ref{Sceq3}) and (\ref{Sceq4}) degenerate into
\be\label{Sceq5}
{\mC}_{1}\zeta'+{\mC}_{2}\zeta=0.
\ee
Unless ${\mC}_{1}={\mC}_{2}=0$ , the constraint (\ref{Sceq5}) states that
there is no scalar dynamical DoF  at the linear perturbation level.
This is a bit strange because the action (\ref{model0}) clearly shows that there is at least one scalar dynamical DoF contributed by $\phi$.
This contradiction implies that the model (\ref{model0}) suffers from strong coupling issue in  flat universe when $f_{3}=f_{4}$ but ${\mC}_{1}\neq 0$ or ${\mC}_{2}\neq0$.
In addition, the absence of scalar DoF also makes the inflaton $\phi$ unable to provide the primordial density perturbation.
To overcome these difficulties, we need to additionally require ${\mC}_{1}={\mC}_{2}=0$ when $f_{3}=f_{4}$.

To sum up, after satisfying the ghost-free condition $f_{24}=0$, we still need to require the following conditions to avoid strong coupling problem on the background with $\phi'\neq0$,
\be\label{goodf02}
f_{3}\neq f_{4}\quad \text{or} \quad f_{3}-f_{4}={\mC}_{1}={\mC}_{2}=0.
\ee
Eq.~(\ref{goodf02}) means that either $f_{3}\neq f_{4}$ or $f_{3}-f_{4}={\mC}_{1}={\mC}_{2}=0$ is satisfied.
After that, the quadratic action for scalar perturbations can be reduced to the action (\ref{Saction5}),
and the action (\ref{Saction5}) shows that there is one healthy dynamical DoF.

\subsubsection{The case of $f_{24}=0$ on the background with $\phi'=0$}\label{Sperturbation03}

Note that neither the unitary gauge
nor the gauge invariant $\zeta$ is well-defined on the background with $\bar{\phi}'=0$.
In order to make the discussion cover the cases of de Sitter and Minkowski background, we introduce two new variables as
\be\label{newinvariant}
\xi=\delta\phi+\frac{\phi'}{\mH}\psi,~~\sigma=\psi+\mH\beta.
\ee
These two variables are well-defined on background with $\bar{\phi}'=0$.
Then we can express the action (\ref{Saction4}) in terms of $\xi$, $\sigma$ and $\tl$ as
\bea\label{Saction6}
S^{(2)}_{S}=\int \zd\eta\, \zd^{3}k\, a^{2} \left[
\frac{1}{2}(\xi^{'2}-m_{\xi}^{2}\xi^{2})+2k^{2}X\mH^{-1}(f_{3}-f_{4})\tl\sigma
+\hat{\mC}_{1}\xi'\tl+\hat{\mC}_{2}\xi\tl+\hat{\mC}_{3}\tl^{2}
\right],
\eea
where
\bea
& &\nonumber
m_{\xi}^{2}=k^{2}+a^{2}V_{\phi\phi}+\phi'(\phi'V+2\mH V_{\phi})/\mH^{2}\\
& &\nonumber
\hat{\mC}_{1}=-\phi'\mH \hat{\vc}_{2}/a^{2}\\
& &\nonumber
\hat{\mC}_{2}=-(\mH V_{\phi}+\phi'V)\hat{\vc}_{2}-4\mH\hat{\vc}_{3}\\
& &\nonumber
\hat{\mC}_{3}=X^{2}[4\hat{\vc}_{1}-a^{2}V\mH^{-2}(f_{3}-f_{4})](f_{3}-f_{4})\\
& &\nonumber
\hat{\vc}_{1}=\phi'f_{0X}/\mH+2 f_{1X}+3 (f_{3}-f_{4})+3X(f_{3X}-f_{4X})\\
& &\nonumber
\hat{\vc}_{2}=4f_{3}+(2a^{2}V/\mH^{2}-4)(f_{3}-f_{4})-4\hat{\vc}_{1}\\
& &\nonumber
\hat{\vc}_{3}=f_{0}-f_{1\phi}+Xf_{4\phi}+2\big(V_{\phi}+3\phi'\mH/a^{2}\big)(f_{1X}-f_{4}-Xf_{4X})+\phi'\mH(3f_{3}-f_{4})/a^{2}\\
& &\nonumber
\eea
Although the action (\ref{Saction6}) is obtained in the unitary gauge,
the action (\ref{Saction6}) should hold in any gauge that satisfies $\omega^{A}_{~B\mu}=0$,
because variables $\xi$, $\sigma$ and $\tl$ are invariant under the infinitesimal diffeomorphism
\footnote{The gauge transformation law of perturbation variables in Eq.~(\ref{tetradperturbation}) can be found in Sec.~IV.B of Ref.~\cite{Li:2021wij}.
It is easy to verified that the variables $\xi$, $\sigma$ and $\tl$ are invariant under infinitesimal diffeomorphisms.
}.
Therefore, the action (\ref{Saction6}) can be applied to the background with $\bar{\phi}'=0$.

On the background with $\bar{\phi}'=0$, that is, the de Sitter background, the action (\ref{Saction6}) can be reduced to
\bea\label{SactiondS}
S^{(2)}_{S}=\int \zd\eta\, \zd^{3}k\, a^{2}\left\{
\frac{1}{2}\left[\xi^{'2}-(k^{2}+a^{2}V_{\phi\phi})\xi^{2}\right]
+4\mH(f_{1\phi}-f_{0})\xi\tl
\right\}
\eea
The action (\ref{SactiondS}) can also be obtained directly in the Newton gauge (see Appendix \ref{CNgauge} for details).
It can be seen that $\tl$ is non-dynamical field and the variations of the action with respect to it lead to the following constraint
\be\label{CeqdS}
\mH(f_{1\phi}-f_{0})\xi=0.
\ee
As analyzed in the Sec.~\ref{Sperturbation02}, there should be at least one dynamical DoF in scalar perturbations because
the action (\ref{model0}) contains the dynamical term of the scalar field $\phi$.
This requires that the constraint (\ref{CeqdS}) can always be satisfied automatically, that is,
\be
(f_{1\phi}-f_{0})|_{X=0}=0.
\ee
After that, the action (\ref{SactiondS}) is reduced to
\bea\label{SactionMin}
S^{(2)}_{S}=\int \zd\eta\, \zd^{3}k\, a^{2}\left\{
\frac{1}{2}\left[\xi^{'2}-(k^{2}+a^{2}V_{\phi\phi})\xi^{2}\right]
\right\}
\eea
The action (\ref{SactionMin}) shows that there is one healthy dynamical DoF.

The Minkowski background can be viewed as a de Sitter background with an infinite Hubble radius, i.e. $\mH=0$.
On the Minkowski background,
the action (\ref{SactiondS}) can be directly reduced to the action (\ref{SactionMin}) without imposing any additional conditions.
It means that there is always one dynamical DoF on the Minkowski background.
In order to make the number of dynamical DoF to be background-independent,
we need to require that there is also one dynamical DoF on the de Sitter and flat universe background.
This is exactly what we have done above.

\subsection{Quadratic action for vector perturbations}\label{Vperturbation}

For vector perturbations,  we can expand them with the circular polarization bases, such as
\be\label{Vexpand}
B_{i}^{V}(\eta, \vec{x})=\sum_{\zA}\int \frac{\zd^{3}k}{(2\pi)^{\frac{3}{2}}}\,
B_{\zA}(\eta, \vec{k})\, \hat{e}_{i}^{\zA}(\vec{k})\,\ze^{\zi\vec{k}\cdot\vec{x}},
\ee
where the circular polarization bases $\{\hat{e}_{i}^{\zA}(\vec{k}), \zA=\zL,\zR\}$ satisfy the relation
$\epsilon_{ijk}k_{j}\hat{e}_{k}^{\zA}(\vec{k})=\zi k\pA \hat{e}_{i}^{\zA}(\vec{k})$, where $\vp_{\zL}=-1$ and $\vp_{\zR}=1$.
Note that we use the normal letter $\zA$ for the left- and right- hand indices
to distinguish it from the italic letter $A$ used to represent the tetrad indices.
Then the quadratic action for the vector perturbations can be directly obtained as
\bea\label{Vaction1}
S^{(2)}_{V}=\sum_{\zA} \int \zd\eta\, \zd^{3}k\, a^{2}\bigg[
f_{24}(2\lambda_{\zA}'\beta_{\zA}'-\pA k B_{\zA}\beta_{\zA}')-k^{2}(f_{2}b_{\zA}+2f_{25}\beta_{\zA})\lambda_{\zA}
+\pA k (\vc_{2}\lambda_{\zA}^{2}+ \vc_{3}\beta_{\zA}^{2})+\frac{k^{2}}{4}B_{\zA}^{2}
\bigg],~~
\eea
where
\begin{eqnarray}
	\vc_{3}=-\vc_{2}+a^{-2}[a^{2}X(f_{4}+f_{5}-f_{6})]'.\nonumber
\end{eqnarray}
It can be seen that $B_{\zA}$ is non-dynamical field
and the variation of the action (\ref{Vaction1}) with respect to $B_{\zA}$ leads to the following constraint
\be\label{Vceq1}
B_{\zA}=2f_{2}\lambda_{\zA}+\frac{2\pA f_{24}}{k}\beta_{\zA}'.
\ee
One can eliminate $B_{\zA}$ from the action (\ref{Vaction1}) by substituting
the constraint (\ref{Vceq1}) back into it. After that, the quadratic action for vector perturbations can be expressed as
\bea\label{Vaction2}
& &\nonumber
S^{(2)}_{V}=\sum_{\zA} \int \zd\eta\, \zd^{3}k\, a^{2}\bigg[
2f_{24}\lambda_{\zA}'\beta_{\zA}'-f_{24}^{2}{\beta_{\zA}'}^{2}-2p_{\zA}kf_{2}f_{24}\lambda_{\zA}\beta_{\zA}'
\\ & &\quad\quad\quad\quad\quad\quad\quad\quad\quad\quad
-2k^{2}f_{25}\lambda_{\zA}\beta_{\zA}+(\pA k\vc_{2}-k^{2}f_{2})\lambda_{\zA}^{2}+\pA k\vc_{3}\beta_{\zA}^{2}
\bigg].
\eea

In the case of $f_{24}\neq0$,
the mixing terms ${\beta_{\zA}}'\lambda_{\zA}'$ and $-{\beta_{\zA}'}^{2}$ in the action (\ref{Vaction2})
generally indicate the existences of ghost modes.
In order to explicitly see how many dynamical DoFs and how many ghost modes there are,
we redefine the following independent vector perturbation variables through $\beta_{\zA}$ and $\lambda_{\zA}$:
\be\label{newvector}
\hat{\beta}_{\zA}=\beta_{\zA}-\frac{1}{f_{24}}\lambda_{\zA},~~\hat{\lambda}_{\zA}=\frac{1}{f_{24}}\lambda_{\zA}.
\ee
It can be verified that the transformation (\ref{newvector}) is linearly reversible.
Then we can express the action (\ref{Vaction2}) in terms of the new perturbation variables as
\be\label{Vaction4}
S^{(2)}_{V}=\sum_{\zA} \int \zd\eta\, \zd^{3}k\, a^{2}\bigg[
f_{24}^{2}(\hat{\lambda}_{\zA}^{'2}-\hat{\beta}_{\zA}^{'2})+2\mC_{6}\hat{\lambda}_{\zA}\hat{\beta}_{\zA}'
+2\mC_{7}\hat{\lambda}_{\zA}\hat{\beta}_{\zA}+\mC_{8}\hat{\lambda}_{\zA}^{2}+\pA k\vc_{3}\hat{\beta}_{\zA}^{2}
\bigg].
\ee
where
\bea
& &\mC_{6}=f_{24}(f_{24}'-\pA k f_{2}f_{24}),\nonumber\\
&&\mC_{7}=\pA k\vc_{3}-k^{2}f_{24}f_{25},\nonumber\\
 & &\mC_{8}=\pA k(\vc_{3}+f_{24}^{2}\vc_{2})
-k^{2}f_{24}(2f_{25}+f_{2}^{2}f_{24})-a^{-2}\left(a^{2}\mC_{6}\right)'.\nonumber
\eea
The quadratic action (\ref{Vaction4}) shows that all the four components of vector perturbations,
$\hat\beta_{\zA}$ and $\hat\lambda_{\zA}$ with $\zA=\zL,\zR$, are dynamical modes.
It also clearly shows that both components of $\hat{\beta}_{A}$ are ghost modes because their kinetic terms have wrong signs.
Again, this will cause the vacuum instability and the only way for vector perturbations to avoid ghost instability is to keep $f_{24}=0$.

In the case of $f_{24}=0$,  the quadratic action (\ref{Vaction2}) reduces to
\be\label{Vaction3}
S^{(2)}_{V}=\sum_{\zA} \int \zd\eta\, \zd^{3}k\, a^{2}\bigg[
-2k^{2}f_{25}\lambda_{\zA}\beta_{\zA}+(\pA k\vc_{2}-k^{2}f_{2})\lambda_{\zA}^{2}+\pA k\vc_{3}\beta_{\zA}^{2}
\bigg].
\ee
Obviously in this case all vector perturbations are non-dynamical.
The variations of the action (\ref{Vaction3}) with respect to $\beta_{\zA}$ and $\lambda_{\zA}$ lead to $\beta_{\zA}=\lambda_{\zA}=0$.
So naturally there is no ghost instability.

\subsection{Quadratic action for tensor perturbations}\label{Tperturbation}

For tensor perturbations,  we can expand them as follow
\be\label{Texpand}
h^{T}_{ij}(\eta, \vec{x})=\sum_{\zA}\int \frac{\zd^{3}k}{(2\pi)^{\frac{3}{2}}}\, h_{\zA}(\eta, \vec{k})\, \hat{e}_{i j}^{\zA}(\vec{k})\,
\ze^{\zi\vec{k}\cdot\vec{x}},
\ee
where the circular polarization bases $\{\hat{e}_{ij}^{\zA}(\vec{k}), \zA=\zL,\zR\}$ satisfy the relation
$k_{l}\epsilon_{lik}\hat{e}_{j k}^{\zA}(\vec{k})=\zi \pA k \hat{e}_{i j}^{\zA}(\vec{k})$.
Then the quadratic action for the tensor perturbations can be directly obtained as
\be\label{Taction}
S^{(2)}_{T}=\sum_{\zA} \int \zd\eta\, \zd^{3}k\, \frac{a^{2}}{4}
\Big[ {h_{\zA}'}^{2}-\omega^{2}_{\zA}(k)h_{\zA}^{2}\Big],
\ee
where
\be\label{dispersion}
\omega^{2}_{\zA}(k)=k^{2}+2(\phi'f_{0}-f_{1}'+3\mH f_{23})\pA k.
\ee
From the action (\ref{Taction}), one can obtain the equation of motion for GWs as
\be\label{GWeom}
h_{\zA}'' + 2\mH h_{\zA}' + \omega^{2}_{\zA}(k) h_{\zA}=0.
\ee
Firstly, it can be seen from the action (\ref{Taction}) that the tensor perturbations are ghost-free.
Secondly, the modified dispersion relation $\omega_{\zA}^{2}(k)$ is helicity dependent when $\phi'f_{0}-f_{1}'+3\mH f_{23}\neq0$.
It can be seen from Eq.~(\ref{GWeom}) that this causes GWs with different helicities to
have different phase velocities $v_{p}^{\zA}=\omega_{\zA}/k$, i.e., the velocity birefringence.
Considering small coupling $f_{i}$ and slow evolution of scalar field $\phi$, we can expand the expression of  $v_{p}^{\zA}$ as
\begin{eqnarray}\label{phase velocity}
	v_{p}^{\zA} = \frac{\omega_{\zA}}{k} = 1+\frac{\mathcal{\pA M}}{k}+\mathcal{O}(\mathcal{M}^2),
\end{eqnarray}
here we have defined $\mM=\phi'f_{0}-f_{1}'+3\mH f_{23}$.
This is the explicit signal of parity violation in this model.
We can also see that the phase velocity difference become important only at the region of small $k$ (large scales), so this is an infrared effect.
Within the framework of Riemannian geometry,
the slightly complicated PV gravity model will
have both the velocity birefringence phenomenon and the amplitude birefringence phenomenon on GWs \cite{Gao:2019liu,Zhao:2019xmm}.
But within the TG framework,  even if the NYTG model is extended to the complicated model (\ref{model0}),
the GWs still has only the velocity birefringence phenomenon but no amplitude birefringence phenomenon, and it is still the infrared effect.
It seems that to construct PV gravity models with the amplitude birefringence phenomenon on GWs in the TG framework, we need to bring derivatives of torsion into the action.

The expression of phase velocities of tensor perturbations shows that GWs and light propagate with different velocities. This difference can be tightly constrained by the present gravitational wave experiments.
The authors in paper \cite{Wu:2021ndf} found that  the effects of velocity birefringence can be explicitly presented by the modifications in the GW phase. Confronting such modifications with data of GWs events of binary black hole merges observed
by LIGO-Virgo, they also gave an up bound on the velocity birefringence parameter which corresponds to $|2\mM|/a<6.5\times 10^{-42}\, \mathrm{GeV}$ in this paper. One can see that this bound is very tight and shows that there are no significant signals of the velocity birefringence of GWs. But we also should note that this bound only constrains the parameter $\mM/a$ at present universe,
and since $\mM/a$ depends on the evolution of the universe, it may be significant at early universe.

\subsection{Futher  analysis on stability}

From the analysis in Sec.~\ref{Sperturbation} and Sec.~\ref{Vperturbation},
we know that the only way to avoid ghost instability is to keep
\begin{eqnarray}\label{goodf1}
	f_{24}=f_{2}(\phi,X)+Xf_{4}(\phi,X)=0,
\end{eqnarray}
no matter for scalar perturbations or vector perturbations.
In the NYTG model \cite{Li:2020xjt}, $f_{2}=f_{4}=0$,
the condition (\ref{goodf1}) can be satisfied, so the NYTG model is ghost-free.
In the extended NYTG model \cite{Li:2022mti}, $f_{2}\neq0, f_{4}=0$,
the condition (\ref{goodf1}) cannot be satisfied, so the extended NYTG model suffers from the problem of ghost instability.
It can be seen that the ghost-free condition (\ref{goodf1}) is consistent with previous studies.

After the ghost-free condition (\ref{goodf1}) is satisfied, in order to avoid the strong coupling problem,
we need to additionally require the following conditions
\bea
& &\label{goodf2}(f_{1\phi}-f_{0})|_{X=0}=0,\\
& &\label{goodf3}f_{3}\neq f_{4}\quad \text{or} \quad f_{3}-f_{4}={\mC}_{1}={\mC}_{2}=0,
\eea
These conditions ensure that the number of dynamical DoFs on the Minkowski, de Sitter and flat universe background are the same at the linear perturbation level.
This is necessary to avoid strong coupling.
In addition, these conditions also ensure that there is a healthy dynamical DoF in scalar perturbations,
which can generate the primordial density perturbation through inflation.

From Eq.~(\ref{goodf3}), it seems that there are two feasible conditions.
However, the condition $f_{3}-f_{4}={\mC}_{1}={\mC}_{2}=f_{24}=0$ would give the following requirement
\bea
& &\label{goodf31}
f_{2}+Xf_{3}=f_{2}+Xf_{4}=0,
\\ & &\label{goodf32}
\phi'\mH^{-1}f_{0X}+2f_{1X}-f_{3}=0,
\\ & &\label{goodf33}
\phi'(f_{0}-f_{1\phi}-f_{2\phi})-X'(f_{1X}+f_{2X})+2X\mH f_{3}=0.
\eea
In order to make Eqs.~(\ref{goodf31})-(\ref{goodf33}) independent of the specific background evolution,
that is, independent of the specific values of $\mH$, $X$, etc.,
the functional forms of $\{f_{\vi}(\phi,X),~\vi=0,1...6\}$ can only satisfy the following conditions
\be\label{goodf4}
f_{2}(\phi,X)=f_{3}(\phi,X)=f_{4}(\phi,X)=0,~f_{0}(\phi,X)=f_{0}(\phi)
,~f_{1}(\phi,X)=f_{1}(\phi),~f_{0}(\phi)=\frac{\partial f_{1}(\phi)}{\partial \phi}.
\ee
Note that due to the identity $\mathring{\nabla}_{\mu}\tilde{T}^{\mu}=\mP_{1}$,
when the condition (\ref{goodf4}) is satisfied,
the contribution of $\mP_{0}$ and the contribution of $\mP_{1}$ in the action (\ref{model0}) will cancel out,
so the condition (\ref{goodf4}) is  equivalent to the following condition
\be\label{goodf5}
f_{0}(\phi,X)=f_{1}(\phi,X)=f_{2}(\phi,X)=f_{3}(\phi,X)=f_{4}(\phi,X)=0.
\ee
But the condition (\ref{goodf5}) will make $\omega_{\zA}^{2}(k)=k^{2}$, that is, the dispersion relation of GWs no longer depends on helicity.
This means that GWs have no PV effect at the linear perturbation level and such a property is undesirable for a PV gravity model.
Therefore, in order to preserve the PV signal in GWs, we should adopt $f_{3}\neq f_{4}$ in Eq.~(\ref{goodf3}).

To sum up, if we want to extract a suitable PV gravity model from the action (\ref{model0}),
we should require $\{f_{\vi}(\phi,X),~\vi=0,1...6\}$ to satisfy the following conditions
\be\label{goodf}
f_{2}(\phi,X)=-X f_{4}(\phi,X),~~f_{0}(\phi,0)=\frac{\partial f_{1}(\phi,0)}{\partial \phi},~~f_{3}(\phi,X)\neq f_{4}(\phi,X).
\ee
In this way, not only the ghost instability can be avoided, the number of dynamical DoFs is background-independent,
but also GWs have PV effect at the linear perturbation level.

It should be noted that there may be some DoFs hidden under the cosmological background at the linear perturbation level,
 as in the case of the $f(\mathbb{T})$ model \cite{Ong:2013qja,Golovnev:2020zpv,BeltranJimenez:2020fvy,Hu:2023juh}.
Therefore, we cannot conclude that the model satisfying condition (\ref{goodf}) must avoid the strong coupling problem.
For a healthy PV model, the condition (\ref{goodf}) is a necessary but not necessarily sufficient condition.
To completely solve the strong coupling problem, the study of higher-order perturbations or even Hamiltonian analysis is required.
These analyzes will be left to subsequent studies.

\section{Conclusion}\label{Conclusion}

In this paper, we wrote down all independent parity-odd terms which are at most quadratic in torsion tensor and contain arbitrary first-order derivatives of scalar field in the teleparallel gravity framework. We constructed the general parity violating gravity model using those parity-odd terms and obtained its equations of motions. In preliminary cosmological exploration, we find that all parity-odd terms have no effect on the flat universe background. To probe the parity violating signals in flat universe, we need to consider the cosmological perturbations. In further cosmological application, we considered a submodel of the general parity violating model, which is linear in parity-odd terms coupled with some arbitrary functions of scalar field and its first-order derivatives. We carefully studied its cosmological linear perturbations and analyzed their stability. For tensor modes, we found that they are always ghost-free and generally exhibit a parity violating signal of velocity birefringence. For vector perturbations, we found that in general there are four dynamical degrees of freedom and two of them are ghost modes. The only way to avoid the ghost modes is to make functions satisfy $f_{24}=0$. Similarly, scalar perturbations generally have three dynamical degrees of freedom and one of them is ghost mode, unless $f_{24}=0$. After the ghost-free condition $f_{24}=0$ is satisfied, in order to preserve the dynamics of the scalar perturbations and the parity violating signal of gravitational waves at linear order, we need to further require functions to satisfy $(f_{1\phi}-f_{0})|_{X=0}=0$ and $f_{3}\neq f_{4}$. Finally, we conclude that the submodel with $f_{24}=(f_{1\phi}-f_{0})|_{X=0}=0$ and $f_{3}\neq f_{4}$ is a potentially viable parity violating model in teleparallel gravity, which is ghost free and exhibits the velocity birefringence phenomenon of gravitational waves at linear order.

\subsection*{Acknowledgements}
This work is supported in part by NSFC under Grant No. 12075231 and No. 12247103, and by National Key Research and Development Program of China Grant No. 2021YFC2203102.

\appendix

\section{Proof of some identities}\label{Appindentityproof}

The identities in Eq.~(\ref{PVrelation2}) play a key role in counting the number of independent PV quadratic torsion terms.
Therefore, they are important in this paper and deserve a detailed proof.
In this appendix, we give proofs of these identities.

Without loss of generality, we assume that $\phi^{\mu}$ is timelike and future-directed.
Then for any point $O$ in spacetime, we can choose the coordinate such that
so that $g_{\mu\nu}=\eta_{\mu\nu}$ and $\phi^{\mu}=(\sqrt{X},0,0,0)$ at point $O$.
At point $O$, we denote
\be\label{appB1}
x_{0}=\e_{ijk}\vt_{i00}\vt_{0jk},~~y_{0}=\e_{ijk}\vt_{0il}\vt_{ljk}.
\ee
From $\vt_{[\mu\nu]\rho}=\vt_{(\mu\nu\rho)}=0$, we can obtain
\be\label{appB2}
\vt_{00i}=-2\vt_{i00},~~\vt_{li0}=-(\vt_{0il}+\vt_{0li}).
\ee
Since $\vt_{\mu\nu\rho}$ is completely traceless, we have
\be\label{appB3}
\vt_{\mu i i}=\vt_{\mu00},~~\vt_{i i \mu}=\vt_{00\mu}.
\ee
Also note that
\bea\label{appB4}
\nonumber
2\e_{ijk}\vt_{0[il]}\vt_{ljk}&=&\e_{ijk}\e_{iln}\e_{npq}\vt_{0pq}\vt_{ljk}
=(\delta_{jl}\delta_{kn}-\delta_{kl}\delta_{jn})\e_{npq}\vt_{0pq}\vt_{ljk}
\\
&=&\e_{ijk}(\vt_{lli}-\vt_{ill})\vt_{0jk}=\e_{ijk}(\vt_{00i}-\vt_{i00})\vt_{0jk}=3x_{0}.
\eea
Combining Eq.~(\ref{appB1}), (\ref{appB2}) and (\ref{appB4}), we can obtain
\be\label{appB5}
\e_{ijk}\vt_{0li}\vt_{ljk}=3x_{0}+y_{0},~~ \e_{ijk}\vt_{il0}\vt_{ljk}=-(3x_{0}+2y_{0}).
\ee
Next we can expand $P_{2}$, $J_{4}$, $J_{5}$, $J_{6}$ and $J_{7}$ into coordinate components,
and use Eqs.~(\ref{appB1})-(\ref{appB5}) to simplify these results to
\be\label{appB0}
P_{2}=-6(3x_{0}+y_{0}),~J_{4}=6Xx_{0},~J_{5}=-X(5x_{0}+y_{0}),~
J_{6}=X(x_{0}-y_{0}),~J_{7}=X^{2}x_{0}.
\ee
Through Eq.~(\ref{appB0}), we can simply verify that the identities in Eq.~(\ref{PVrelation2}) do hold at point $O$.
Although Eq.~(\ref{appB0}) is coordinate-dependent, the identities in Eq.~(\ref{PVrelation2}) are coordinate-independent.
And because $O$ can be any point in spacetime, the identities in Eq.~(\ref{PVrelation2}) hold in the whole spacetime.

\section{Scalar perturbations on de Sitter background in Newton gauge}\label{CNgauge}

In Sec.~\ref{Sperturbation03}, we obtain the quadratic action (\ref{SactiondS}) for scalar perturbations on the de Sitter background
by means of the quadratic action (\ref{Saction6}) on the flat universe.
Since the action (\ref{Saction6}) is obtained in the unitary gauge which is ill-defined on the de Sitter background,
one might question the reliability of the action (\ref{SactiondS}).
To ensure that the action (\ref{SactiondS}) is reliable, in this appendix,
we will calculate the quadratic action for scalar perturbations on the de Sitter background in the Newton gauge.

The so-called Newton gauge is to set the perturbation variables $\alpha$ and $B$ in Eqs.~(\ref{tetradperturbation}) and (\ref{metricperturbation}) to zero.
In contrast to the unitary gauge, the Newton gauge is well-defined on the background with $\bar{\phi}'=0$,
which makes the results in the Newton gauge more plausible when we consider the de Sitter background.

In the Newton gauge, the quadratic action for scalar perturbations on the de Sitter background
can be obtained directly from the action (\ref{model0}) as
\bea\label{SactiondS1}
& &\nonumber
S^{(2)}_{S}=\int \zd\eta\, \zd^{3}k\, a^{2}\bigg\{
\frac{1}{2}\left[\delta\phi^{'2}-(k^{2}+a^{2}V_{\phi\phi})\delta\phi^{2}\right]
-3\psi^{'2}-a^{2}V A^{2}
\\ & &\quad\quad\quad\quad\quad\quad\quad\quad
-(6\mH\psi'+2k^{2}\psi)A+k^{2}\psi^{2}+4\mH(f_{1\phi}-f_{0})\delta\phi\tl
\bigg\},
\eea
where we have imposed the ghost-free condition $f_{24}=0$.
Obviously $A$ is non-dynamical
and the variations of the action (\ref{SactiondS1}) with $A$ lead to the following constraint
\be\label{CeqdS1}
A=-\frac{1}{a^{2}V}(3\mH\psi'+k^{2}\psi^{2}).
\ee
Substituting Eq.~(\ref{CeqdS1}) back into the action (\ref{SactiondS1}), then the action (\ref{SactiondS1}) can be reduced to
\bea\label{SactiondS2}
S^{(2)}_{S}=\int \zd\eta\, \zd^{3}k\, a^{2}\bigg\{
\frac{1}{2}\left[\delta\phi^{'2}-(k^{2}+a^{2}V_{\phi\phi})\delta\phi^{2}\right]
+\frac{k^{4}}{a^{2}V}\psi^{2}
+4\mH(f_{1\phi}-f_{0})\delta\phi\tl
\bigg\}.
\eea
It can be seen that the kinetic term of $\psi$ is canceled out in this process.
So $\psi$ is non-dynamical and the variations of the action (\ref{SactiondS2}) with $\psi$ lead to the following constraint
\be\label{CeqdS2}
\psi=0.
\ee
Substituting Eq.~(\ref{CeqdS2}) back into the action (\ref{SactiondS2}), then the action (\ref{SactiondS2}) can be reduced to
\bea\label{SactiondS3}
S^{(2)}_{S}=\int \zd\eta\, \zd^{3}k\, a^{2}\bigg\{
\frac{1}{2}\left[\delta\phi^{'2}-(k^{2}+a^{2}V_{\phi\phi})\delta\phi^{2}\right]
+4\mH(f_{1\phi}-f_{0})\delta\phi\tl
\bigg\}.
\eea
This is exactly the action (\ref{SactiondS}) because $\xi=\delta\phi$ on the de Sitter background.
The fact that two different methods give the same result demonstrates that the action (\ref{SactiondS}) is reliable.

{}

\end{document}